\documentclass[aps,12pt,preprintnumbers,nofootinbib,superscriptaddress]{revtex4}
\usepackage{graphicx}
\usepackage{amssymb,amsmath}
\usepackage{slashed}
\usepackage{dcolumn}
\begin{document}

\title{\Large{\bf Non-local correction to the energy-momentum tensor for $\phi^{3}$ theory in six dimensions} }

\author{Feng Wu}
\email[Electronic address: ]{fengwu@ncu.edu.cn}
\affiliation{%
Department of Physics, Nanchang University,
330031, China}

\date{\today}

\begin{abstract}
Applying the background field method, we construct by explicit computation the leading-order non-local quantum correction to the on-shell effective action for $\phi^3$ theory in six dimensions. We then use the resulting action to obtain the non-local correction to the energy-momentum tensor. At leading order, we find that this non-local correction modifies the virial current when the scalar field is minimally coupled to gravity. This is to be compared to the classically Weyl invariant case, where it only corrects the traceless part of the energy-momentum tensor. 
\end{abstract}
\maketitle 
\newpage

\date{\today}

\section{Introduction}
Energy and momentum are two very important quantities in physics. In most field theories, the densities of these two quantities can be concisely described by one object, the energy-momentum tensor; it plays a crucial role associated with the global symmetries of a theory.

To study the energy-momentum tensor, it is convenient to consider a quantum field theory of interest in a general curved background since then the energy-momentum tensor is naturally defined as the composite field that is conjugate to the background metric. Specifically, the symmetric (Belinfante) energy-momentum tensor $T_{\mu\nu}$ is defined to be the metric variation of the action 
\begin{equation}
T_{\mu\nu} \equiv {2\over \sqrt{\vert g\vert}}{\delta S \over \delta g^{\mu\nu}}. \label{em} 
\end{equation} 
Properties of the energy-momentum tensor can then be obtained by taking the flat space limit of the field theory of interest.

The energy-momentum tensor $T_{\mu\nu}$ can be decomposed into two parts: the trace part $T_{\mu}^{\mu}$ and the symmetric traceless part. The trace part $T_{\mu}^{\mu}$ is especially important in quantum field theory considering that the advance of the dynamics of a quantum field theory between the end points of the renormalization group flow under scaling is controlled by $n$-point correlation functions of $T_{\mu}^{\mu}$ \cite{Polchinski}. In the case where a field theory is classically scale-invariant, the classical trace $T_{\mu}^{\mu}$ is essentially given by 
\begin{equation}
T_{\mu}^{\mu} = \partial_{\mu} V^{\mu}, \label{virial} 
\end{equation} 
where the possible non-vanishing $V^{\mu}$ is the ``virial current". If one can further show that the virial current is equal to a total derivative: $V^{\mu}(x)=\partial_{\nu} L^{\mu\nu}(x)$, then the scale-invariant theory is also invariant under special conformal transformations and the energy-momentum tensor can be ``improved" to be traceless \cite{Callan}. It is believed that a Poincar$\acute{\rm e}$-invariant interacting field theory that is scale-invariant but not conformally invariant must be non-unitary. However, the verity of the above statement has only been shown in two and four dimensions\footnote{We should note that the 4d case has only been proven in perturbation theory.} \cite{Polchinski, Luty, Dymarsky, Dymarsky2}.

In the quantum theory, the energy-momentum tensor must be regularized and an anomaly arises. In general, the trace of the energy-momentum tensor $T_{\mu}^{\mu}$ contains terms involving the curvature and the terms contributing even in the flat space limit. The curvature-dependent terms have received much attention since Cardy's conjecture \cite{Cardy} that the integral of the trace $T_{\mu}^{\mu}$ in even dimensions $d$ over the sphere $S^{d}$
\begin{equation}
\int_{S^d}\sqrt{\vert g\vert} \langle T_{\mu}^{\mu} \rangle d^{d} x
\end{equation} 
is a monotonically decreasing object along the renormalization group flow. This presumably puts restrictions on the low energy limit of the theory. So far, this proposal has been shown to be correct in two dimensions \cite{Zamolodchikov}. In the four dimensional case, a perturbative proof is given by analyzing the local renormalization group \cite{Jack}. Outside of perturbation theory, only a weak version of the conjecture has been shown \cite{Komargodski}. However, a challenge to this proposal in six dimensions has been published recently in \cite{Grinstein}. 

In this paper, we will by contrast focus on the part of the energy-momentum tensor that contributes in the flat space limit. In flat spacetime, the trace anomaly of the energy-momentum tensor is given by 
\begin{equation}
T_{\mu}^{\mu} = B^{I} {\partial\mathcal{L} \over \partial \lambda^{I}} \label{Beta} , 
\end{equation} 
where $B^{I}$ is the scheme independent beta function of the renormalized coupling constants $\lambda^{I}$ and $\mathcal{L}$ denotes the renormalized Lagrangian of the theory. Note that when there is more than one kind of matter field, the beta function for each coupling constant will in general be renormalization scheme dependent. However, the Ward identity for the flavour current allows one to express the trace anomaly~(\ref{Beta}) in an unambiguous way. For more details on this issue, we refer to \cite{Luty, Baume}.

The quantum corrections to the energy-momentum tensor itself are not as well-known as those to the trace part. For an interacting theory with massless particles, quantum effects give rise to both local and non-local contributions to the effective action. This occurs naturally since separated particles in spacetime are always influenced by each other's long-range forces. The non-local effective action can be comprehended as the result of the renormalization group flows caused by integrating out the high-momentum massless degrees of freedom. This non-local quantum effective action induced by the long-range interactions mediated by the massless particles will in turn give rise to a non-local contribution the the energy-momentum tensor. Although the study of the energy-momentum tensor is an old topic, less attention has been drawn to this point.

In this paper, we investigate $\phi^3$ theory in six dimensions and present the non-local contribution to the energy-momentum tensor, to the first nontrivial order in the $\phi^3$ coupling $\lambda$. Our work is motivated by a recent work by Donoghue and El-Menoufi \cite{Donoghue}, in which the idea originating from the study of gravitational conformal anomalies \cite{Deser} was used to derive the non-local effective action for scalar QED and several aspects of the low energy physics were explored. The $\phi^3$ theory in six dimensions is in a sense one of the simplest classically scale-invariant field theories with marginal interactions and the study of it has a long history. We refer to \cite{Osborn, Grinstein2} and references therein for issues related to the local part of the effective action for this theory and focus our attention on the non-local correction from quantum effects. Following the procedure laid out in \cite{Donoghue}, we construct the quantum effective action for $\phi^3$ theory within the context of perturbation theory, and show by explicit computation that while the correction to the trace of the energy-momentum tensor is local and of order $\lambda^3$, the leading order non-local energy-momentum tensor is of order $\lambda^2$. To our knowledge, these results have not been presented in the literature, and one goal of this work is intended to fill this gap. 
      
The rest of the paper is organized as follows. In Sec. II, some properties of $\phi^3$ theory in six dimensions related to Weyl symmetry are described. Most of the remarks in this section are well-known. Then, as a warm-up exercise, we follow the unconventional method of Ref. \cite{Donoghue} to derive the trace anomaly from the variation of the non-local effective action under scaling in flat space. In Sec. III, we put $\phi^3$ theory on a curved background and use the background field method to construct the quantum effective action. We then used the resulting action to obtain the leading order non-local part of the the energy-momentum tensor. Related issues are discussed along the way. Our conclusions are given in the final section. Useful integrals needed for our computations are listed in appendix A.  

\section{Trace anomaly}
We start with the action for a real scalar field $\phi$ minimally coupled to gravity in six dimensional spacetime with metric $g_{\mu\nu}$ with a marginal self-interaction:
\begin{equation}
S_{\min}=\int d^6 x \sqrt{\vert g \vert} ({1\over 2} g^{\mu\nu} \partial_{\mu} \phi \partial_{\nu} \phi - {1\over 3!}\lambda \phi^3),  \label{Smin}
\end{equation}
where $\lambda$ is the dimensionless coupling constant for the marginal interaction. In the flat space limit, the action~(\ref{Smin}) describes a classically scale-invariant field theory. Indeed, it is obvious to see that the action~(\ref{Smin}) is invariant under the scale transformations defined by
\begin{equation}
x \longrightarrow \eta x,\,\,\, \,\,\,\, \phi \longrightarrow \eta^{-2} \phi. \label{scale}
\end{equation}
Classically, this theory is also invariant under special conformal transformations, and thus is a conformal field theory. A simple way to realize this is using the definition~(\ref{em}) to compute the trace part of the energy-momentum tensor. The result is
\begin{equation}
T_{\mu}^{\mu}=-2\partial_{\mu}\phi \partial^{\mu} \phi + \lambda \phi^3 = -\Box \phi^2 \label{ftrace},
\end{equation}
 where $\Box$ is the Laplace operator in six dimensions and we have used the equation of motion for the field $\phi(x)$ to obtain the last expression. As is well-known and stated in the previous section, the fact that the virial current contained in~(\ref{ftrace}) is a total derivative indicates that the energy-momentum tensor can be ``improved" to be traceless by modifying the Lagrangian in~({\ref{Smin}). In this case, it is achieved by adding to the minimally coupled Lagrangian a term of the form $R\phi^2$. 
 
 The reason for the non-vanishing result~(\ref{ftrace}) is due to the fact that the minimally coupled action~(\ref{Smin}) is not invariant under the local Weyl transformations defined by
 \begin{equation}
 g_{\mu\nu} \longrightarrow e^{2\tau (x)}g_{\mu\nu} ,\,\,\, \,\,\,\, \phi(x)\longrightarrow e^{-2\tau(x)} \phi.
 \end{equation}
 Instead, the modified action\footnote{Marginal interactions $R^2\phi$ and $R\nabla^2 \phi$ are irrelevant to the discussion in this paper and are therefore neglected.} defined by
 \begin{equation}
 S_{\xi}=S_{min}+\xi \int d^6 x \sqrt{\vert g \vert}R\phi^2  \label{Sxi}
 \end{equation}
 is invariant under Weyl transformations with the choice $\xi={1\over10}$, and thus the trace of the energy-momentum tensor vanishes in this case. In fact, in terms of the manifestly Weyl invariant metric defined by
 \begin{equation}
 \hat{g}_{\mu\nu} \equiv \phi g_{\mu\nu},
 \end{equation}
 the modified action $S_{\xi}$ with $\xi={1\over 10}$ can be rewritten as the Weyl$\times$diffeomorphism invariant Einstein-Hilbert action for $\hat{g}$:
 \begin{equation}
 S_{\xi={1\over10}}={1\over 10}\int d^6 x \sqrt{\vert \hat{g} \vert} (\hat{R} + 2\Lambda), 
 \end{equation}
 where $\Lambda \equiv -{5\over 6} \lambda$ is the dimensionless ``cosmological constant"  .
 
 We now proceed to derive the trace anomaly for $\phi^3$ theory in flat space by using the uncustomary method elaborated in \cite{Donoghue}. It starts with the computation of the effective action, denoted by $\Gamma$. After obtaining $\Gamma$, the trace anomaly follows directly from the variation of the non-local $\Gamma$ under scaling. 
 
 We begin by splitting the scalar field $\phi$ into a classical part $\phi_{c}$ and a quantum fluctuation $\zeta$: 
\begin{equation}
\phi(x)=\phi_{c}(x)+\zeta(x). \label{phisplit}
\end{equation}
Expanding the Lagrangian $\mathcal{L}$ for $\phi^3$ theory in flat space about $\phi_{c}$ and keeping only the terms up to quadratic order in $\zeta$ for our purposes, we have
 \begin{equation}
 \mathcal{L}[\phi]={1\over 2} \partial_{\mu} \phi_{c} \partial^{\mu}\phi_{c} -{1\over 3!} \lambda \phi_{c}^3 +{1\over 2} \partial_{\mu} \zeta \partial^{\mu} \zeta - {1\over 2} \lambda\phi_{c} \zeta^2.   \label{L1}
 \end{equation}
Note that we have dropped terms linear in $\zeta$ by applying the classical field equation. According to the background field method, the lowest-order quantum effective action denoted $\Gamma^{(1)}$, obtained by integrating out $\zeta$, is given by
\begin{equation}
\Gamma^{(1)}[\phi_{c}] = \int d^6 x \mathcal{L} [\phi_{c}] +{i\over 2} {\rm ln \,\,det}[-{\delta^2 \mathcal{L} \over \delta\phi\delta\phi}] +\int d^6 x \mathcal{\delta L}, \label{eff}
\end{equation}
where $\delta\mathcal{L}$ contains all the counterterms to be determined by the renormalization conditions. 

From the terms quadratic in $\zeta$ in~(\ref{L1}), we have
\begin{equation}
-{\delta^2 \mathcal{L} \over \delta\phi\delta\phi}=\Box + \lambda \phi_{c}.
\end{equation}
After expanding the functional determinant in powers of the coupling $\lambda$ and neglecting the irrelevant constant independent of $\lambda$ and $\phi_{c}$, we have
\begin{equation}
\Gamma^{(1)} [\phi_{c}]=\int d^{6} x (\mathcal{ L} [\phi_{c}]+\mathcal{\delta L})+{i\over 2}{\rm Tr} [ {1\over \Box} \lambda \phi_{c} -{1\over 2} {1\over \Box} \lambda \phi_{c} {1\over \Box} \lambda \phi_{c}+{1\over 3} {1\over \Box} \lambda \phi_{c} {1\over \Box} \lambda \phi_{c} {1\over \Box} \lambda \phi_{c}+....].\label{eff2}
\end{equation}
To compute the terms in the expansion in~(\ref{eff2}), we adopt dimensional regularization to evaluate the integrals in $d=6-\epsilon$ dimensional spacetime. In this scheme, the massless tadpole graph, corresponding to the first term in the trace in~(\ref{eff2}), can be taken to vanish. The first two contributing terms correspond to the Feynman diagrams shown in Fig. 1.

At this point, it should be noted that in principle the effective action can contain completely new interactions. However, it is known that the trace anomaly arises from the subtraction of divergences from quantum effects and thus only divergent diagrams contribute it. At leading order, such diagrams are nothing but the ones shown in Fig. 1. 
\begin{figure}[t]
\begin{center}
\includegraphics[width=10cm,clip=true,keepaspectratio=true]{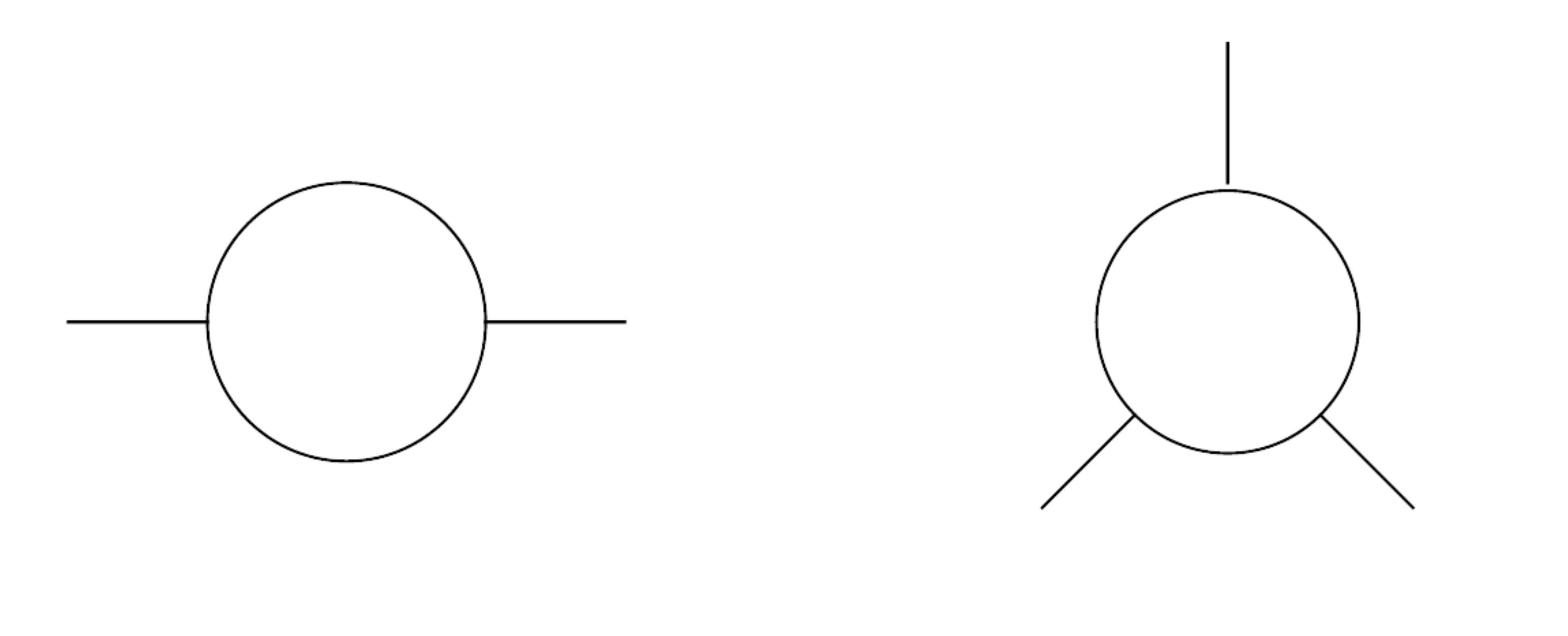}
\caption{\small Leading order Feynman diagrams corresponding to the divergent terms in the expansion of the functional determinant in~(\ref{eff2}).}
\end{center}\label{loop1}
\end{figure}

We now compute the divergent terms in the expansion of the functional determinant explicitly. The term of order $\lambda^2$ is
\begin{equation}
-{1\over 2} {\rm Tr}[{1\over \Box} \lambda \phi_{c} {1\over \Box} \lambda \phi_{c}]= -{\lambda^2 \over 2} \int {d^6 k \over (2\pi)^6} \tilde{\phi}_{c}(k)\tilde{\phi}_{c}(-k) I_{2}(k), 
\end{equation}
where $\tilde{\phi}_{c}(k)$ is the Fourier transform of $\phi_{c}(x)$ and the bubble integral $I_{2}(k)$ is given by
\begin{equation}
I_{2}(k)=\int {d^6 l \over (2\pi)^6 } {1\over l^2 (k+l)^2}.
\end{equation}
Using the formula~(\ref{buint}) shown in Appendix A, we obtain
 \begin{equation}
-{1\over 2} {\rm Tr}[{1\over \Box} \lambda \phi_{c} {1\over \Box} \lambda \phi_{c}]= -{i\over 12 (4\pi)^3} \lambda^2 \int {d^6 k \over (2\pi)^6} \tilde{\phi}_{c}(k)\tilde{\phi}_{c}(-k)k^2\left({1\over \bar{\epsilon}} +{8\over 3} - {\rm ln} (-k^2) \right), \label{b}
\end{equation}
where ${1\over \bar{\epsilon}} \equiv {2\over \epsilon} -\gamma +{\rm ln} 4\pi$.

The term of order $\lambda^3$ is
 \begin{equation}
{1\over 3} {\rm Tr}[{1\over \Box} \lambda \phi_{c} {1\over \Box} \lambda \phi_{c}{1\over \Box} \lambda\phi_{c}]= -{\lambda^3\over 3} \int {d^6 p_{1}\over (2\pi)^6} {d^6 p_{2}\over (2\pi)^6}\tilde{\phi}_{c}(p_{1})\tilde{\phi}_{c}(p_{2}) \tilde{\phi}_{c}(p_{3})I_{3}(p_{1},p_{2}) ,
\end{equation}
where $p_{1}^{\mu}+p_{2}^{\mu}+p_{3}^{\mu}=0$ and the triangle integral $I(p_{1},p_{2})$ is defined by
\begin{equation}
I_{3}(p_{1},p_{2})=\int {d^6 l \over (2\pi)^6 } {1\over l^2 (l+p_{1})^2(l-p_{2})^2}.
\end{equation}
Applying the formula~(\ref{tri}), we have
\begin{equation}
{1\over 3} {\rm Tr}[{1\over \Box} \lambda \phi_{c} {1\over \Box} \lambda \phi_{c}{1\over \Box} \lambda\phi_{c}]= {i\over 6(4\pi)^3}\lambda^3 \int {d^6 p_{1}\over (2\pi)^6} {d^6 p_{2}\over (2\pi)^6}\tilde{\phi}_{c}(p_{1})\tilde{\phi}_{c}(p_{2}) \tilde{\phi}_{c}(p_{3})\left({1\over \bar{\epsilon}} -{\rm ln}(-s_{12}) +{\rm finite} \right), \label{t}
\end{equation}
where we have defined the Mandelstan variable $s_{12}\equiv (p_1+p_2)^2$. Note that the finite part in~(\ref{t}) is invariant under scaling. This fact can be seen clearly from its explict form in~(\ref{finite1}) and~(\ref{finite2}).

After absorbing the divergences of these two diagrams into the counterterms $\delta\mathcal{L}$ in~(\ref{eff2}) using the modified minimal subtraction scheme, we obtain from the results~(\ref{b}) and~(\ref{t}) the following expression for the renormalized effective action that violates scale symmetry, in position space:
\begin{eqnarray}
\Gamma^{(1)}[\phi_{c}]&=&{\lambda^2 \over 24 (4\pi)^3} \int d^6x d^6y\, \phi_{c}(x) \langle x \vert {\rm ln}\left({\Box \over \mu^2}\right)\vert y \rangle\left( \Box \phi_{c}(y) +2\lambda \phi_{c}^{2}(y)\right)+\left({\rm scale\,\,inv.\,\,part} \right) \nonumber\\
&=& {1\over 16}{\lambda^3 \over (4\pi)^3} \int d^6x\,\phi_{c}(x) {\rm ln}\left({\Box \over \mu^2}\right) \phi_{c}^{2}(x)+\left({\rm scale\,\,inv.\,\,part} \right), \label{reff1}
\end{eqnarray}
where in the last step we have used the equation of motion for the field $\phi_{c}$. Using this non-local result, we can now derive the trace anomaly. Explicitly, under the scale transformation~(\ref{scale}), we see that
\begin{equation}
{\rm ln}\,\Box \longrightarrow {\rm ln}\,\Box - 2\, { {\rm ln}\,\eta},
\end{equation}
and so the renormalized effective action~(\ref{reff1}) transforms as
\begin{equation}
\Gamma^{(1)}[\phi_{c}] \longrightarrow \Gamma^{'(1)}\equiv \int d^6x \mathcal{L}^{'}=\Gamma^{(1)}[\phi_{c}] -{1\over 8}{\lambda^3\over(4\pi)^3}{\rm ln}\,\eta \int d^6 x \phi^{3}(x).
\end{equation}
Notice that the scale transformations~(\ref{scale}) is equivalent to the global Weyl transformations
\begin{equation}
 g_{\mu\nu} \longrightarrow \eta^2 g_{\mu\nu} ,\,\,\, \,\,\,\, \phi(x)\longrightarrow \eta^{-2} \phi(x).
 \end{equation}
Using this fact and the definition~(\ref{em}) of the energy-momentum tensor, it is straightforward to show that in the flat space the anomaly induced by the quantum correction is given by 
\begin{equation}
T_{\mu}^{\mu}=-{\partial \mathcal{L}' \over \partial {\rm ln}\,\,\eta}={1\over 8} \left({\lambda \over 4\pi} \right)^3 \phi^3 = \beta(\lambda) {\partial \mathcal{L}[\phi_{c}]\over \partial \lambda},\label{traceanomaly}
\end{equation}
where $\beta(\lambda)=-{3\over 4}({\lambda\over 4\pi})^3$ is the beta function for the coupling constant $\lambda$. In the last step, we have reexpressed the result in the usual form~(\ref{Beta}).

As pointed out in \cite{Donoghue}, this derivation shows transparently that the trace anomaly does not follow from any local physics and is an infrared effect. The violation of scale invariance has its source in the non-local terms of the effective action, which are in turn due to the long-range interactions mediated by the massless degrees of freedom in loops. 

\section{Non-local Energy-momentum Tensor}
When using a local Lagrangian in the action functional as a starting point of a quantum field theory, we have presuppose the locality of the classical energy-momentum tensor of the theory. However, as we have seen in the previous section, quantum effects can induce non-local corrections. In particular, we computed the lowest-order non-local effective action for $\phi^3$ theory in flat space and used it to derive the trace anomaly. At this point, it is natural to raise the question: what is the non-local quantum correction to the energy-momentum tensor besides the trace part? The most convenient way to answer this question is to generalize the analysis of the previous section by considering the field theory of interest in the curved background.  In the remainder of this section, we will use this strategy to determine the leading-order non-local correction to the energy-momentum tensor of $\phi^{3}$ theory and show that it is of order $\lambda^2$.

Let us start with the action~(\ref{Sxi}). First, it is useful to expand the metric about the flat metric $\eta_{\mu\nu}$: $g_{\mu\nu}=\eta_{\mu\nu}+h_{\mu\nu}$. The convention for $\eta_{\mu\nu}$ has signature $ (+,-,-,-,-,-)$. Keeping only the terms up to linear order in $h_{\mu\nu}$, the action~(\ref{Sxi}) takes the form 
\begin{equation}
S_{\xi}[\phi]=\int d^6x \{ {1\over 2} \partial_{\rho} \phi \partial^{\rho} \phi - {1\over 3!} \lambda \phi^3 -{1\over2}h^{\mu\nu} \partial_{\mu} \phi \partial_{\nu} \phi +{1\over 2} h ( {1\over 2}  \partial_{\rho} \phi \partial^{\rho} \phi - {1\over 3!} \lambda \phi^3)-\xi(\Box h -\partial_{\mu}\partial_{\nu}h^{\mu\nu} )\phi^2 \},
\end{equation}
where $h\equiv \eta_{\mu\nu}h^{\mu\nu}$. Given the definition~(\ref{em}), the terms linear in $h^{\mu\nu}$ allows us to read the classical energy-momentum tensor denoted by $T_{\mu\nu}^{cl}$ directly:
\begin{equation}
T_{\mu\nu}^{cl}=\partial_{\mu} \phi \partial_{\nu} \phi-\eta_{\mu\nu}( {1\over 2}  \partial_{\rho} \phi \partial^{\rho} \phi - {1\over 3!} \lambda \phi^3)+2\xi(\Box\eta_{\mu\nu}-\partial_{\mu}\partial_{\nu} )\phi^2 .\label{clT}
\end{equation}
We note that in the classically Weyl invariant case, the last term in~(\ref{clT}) corresponds to the improvement term that renders the classical energy-momentum tensor traceless \cite{Callan}.

To derive the lowest-order quantum correction, we again use the background field method as in the previous section and integrate over the fluctuating quantum field $\zeta$ in the presence of the classical background field $\phi_{c}$, defined in~(\ref{phisplit}). After performing the path-integral, the effective action for the field $\phi_{c}$, to one-loop order, is given by
\begin{equation}
\Gamma_{\xi}^{(1)}[\phi_{c}]=S_{\xi}[\phi_{c}]+{i\over 2} {\rm ln}\, {\rm det}[ \Box +v]+\int d^6x \delta\mathcal{L}^{'},\label{effxi}
\end{equation}
where 
\begin{equation}
v=\lambda(\phi_{c}+{1\over 2}h\phi_{c})-(\partial_{\mu}h^{\mu\nu})\partial_{\nu}-h^{\mu\nu}\partial_{\mu}\partial_{\nu}+{1\over 2}((\partial_{\rho} h)\partial^{\rho}+h\Box)+2\xi(\Box h-\partial_{\mu}\partial_{\nu} h^{\mu\nu})
\end{equation}
and $\delta \mathcal{L}^{'}$ is the counterterm Lagrangian. In the following analysis, we will evaluate the functional determinant on-shell.

The Feynman diagrams correspond to the terms linear in $h^{\mu\nu}$ in the expansion of the functional determinant ${\rm ln\, det}[\Box+ v]$, up to order $\lambda^2$, are shown in Fig. 2.  Fig. 2a and Fig. 2b are massless tadpole diagrams and thus vanish in dimensional regularization scheme. The only term of order $\lambda$ left to be evaluated in the functional determinant is the one corresponding to the diagram shown in Fig. 2c. It reads
\begin{eqnarray}
&-&{\rm Tr} \left[{1\over\Box} \lambda\phi_{c} {1\over \Box}\left(  -(\partial_{\mu}h^{\mu\nu})\partial_{\nu}-h^{\mu\nu}\partial_{\mu}\partial_{\nu}+{1\over 2}((\partial_{\rho} h)\partial^{\rho}+h\Box)+2\xi(\Box h-\partial_{\mu}\partial_{\nu} h^{\mu\nu})
 \right) \right]\nonumber\\
 &=& -\lambda \int {d^6 p \over (2\pi)^{6}} \tilde{\phi}_{c}(-p)\tilde{h}_{\mu\nu}(p)I^{\mu\nu}(p),\label{2c}
\end{eqnarray}   
where $\tilde{h}_{\mu\nu}(p)$ is the Fourier transform of $h_{\mu\nu}(x)$ and we have defined the integral 
\begin{equation}
I^{\mu\nu}(p)=\int {d^6l\over (2\pi)^{6}}{1\over l^2(l+p)^2} \left[ p^{\mu}l^{\nu}+l^{\mu}l^{\nu} -{\eta^{\mu\nu} \over 2}(p\cdot l+l^2)+2\xi(  p^{\mu}p^{\nu}-p^2\eta^{\mu\nu} )   \right].
\end{equation}
Note that the factor $(-1)$ in~(\ref{2c}) is from the expansion and the factor ${1\over 2}$ from the power series of the logarithm cancels the factor 2 from interchanging the vertices. The computation of the integral $I^{\mu\nu}(p)$ by means of standard steps in dimensional regularization gives 
\begin{equation}
I^{\mu\nu}(p)={i\over (4\pi)^{d/2}}\Gamma(2-{d\over2})\left( {\Gamma({d\over 2})^2 \over \Gamma(d)}-2\xi {\Gamma({d\over 2}-1)^2 \over \Gamma(d-2)}   \right) (-p^2)^{{d\over2}-2}(p^2\eta^{\mu\nu}-p^{\mu}p^{\nu})
\end{equation}
which vanishes because of the on-shell condition. Thus the quantum correction to the on-shell effective action is zero up to order $\lambda$. We note in passing that this diagram, in spite of having a positive superficial degree of divergence, is finite for off-shell scalar $\phi_{c}$ in the classically Weyl invariant case where $\xi={1\over 10}$. The fact that for some Feynman integrals the degree of divergence is lower than the one based on power counting because of symmetry is not atypical. For instance, in QED it is known that the one-loop Fermion self-energy is logarithmically rather than linearly divergent because of Lorentz invariance. Here the diagram shown in Fig. 2c is finite at $\xi={1\over 10}$ because of Weyl invariance.  
\begin{figure}[t]
\begin{center}
\includegraphics[width=14cm,clip=true,keepaspectratio=true]{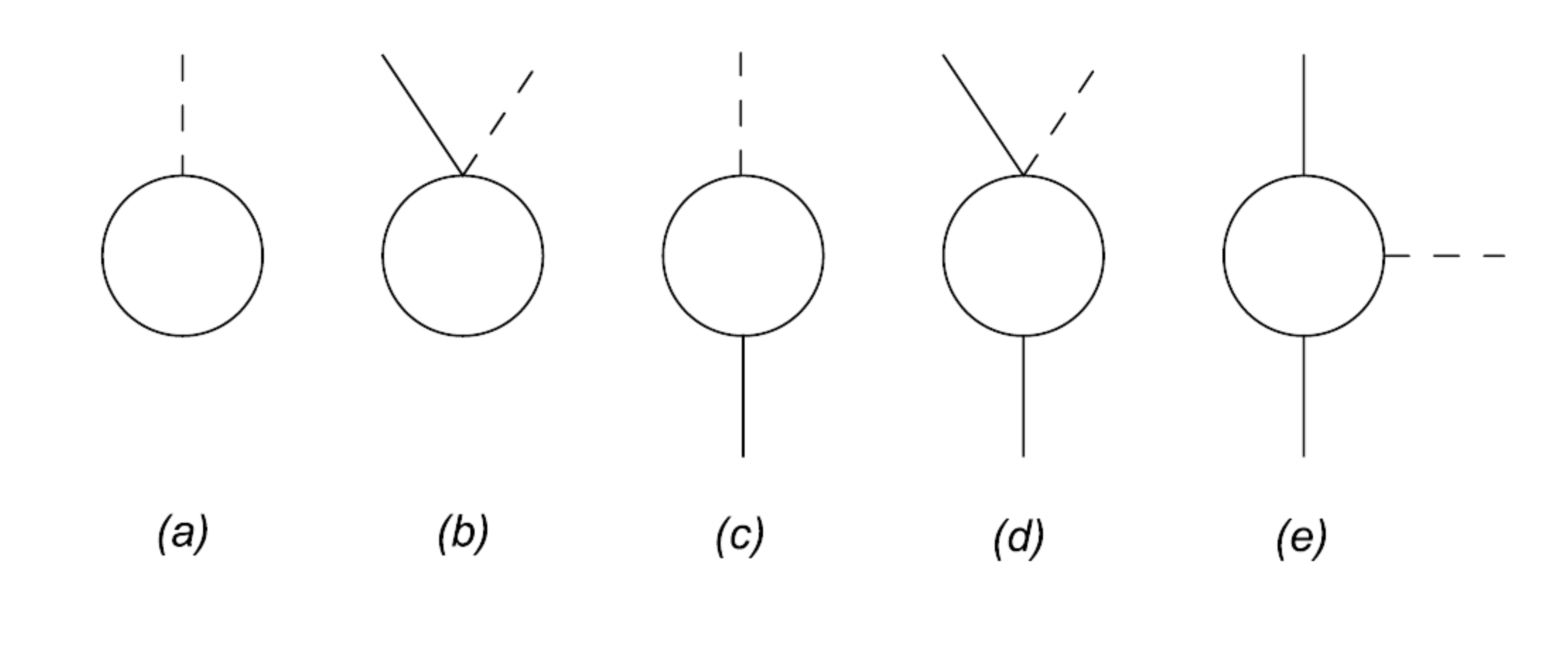}
\caption{\small Feynman diagrams corresponding to the expansion of ${\rm ln\, det}[\Box+ v]$, up to order $\lambda^2$. The external solid, internal solid and dashed lines stand for $\phi_{c}$, $\zeta$ and $h_{\mu\nu}$, respectively.}
\end{center}\label{loop1}
\end{figure}

We next consider the terms of order $\lambda^2$, which correspond to Fig. 2d and Fig.2e. The term in~(\ref{effxi}) which corresponds to Fig. 2d reads
\begin{equation}
-{\rm Tr} \left[  {1\over\Box} \lambda \phi_{c} {1\over \Box} {\lambda\over2} h \phi_{c}\right]=-{\lambda^2\over 2}\int {d^6p\over(2\pi)^{6}}{d^6 p' \over(2\pi)^{6}}\tilde{\phi}_{c}(p')\tilde{\phi}_{c}(-p)\tilde{h}_{\mu\nu}(k) \eta^{\mu\nu} I_{2}(p'),
\end{equation}
where $k^{\mu}=p^{\mu}-p'^{\mu}$ and $I_{2}(p')$ given by~(\ref{buint}). This term vanishes again due to the on-shell condition.

Finally, we will show that the only contribution to the on-shell effective action up to order $\lambda^2$ comes from the diagram shown in Fig. 2e. Explicitly, the term in the functional determinant corresponding to Fig. 2e takes the form
\begin{eqnarray}
&&{\rm Tr}\left[{1\over\Box} \lambda\phi_{c}{1\over\Box} \lambda\phi_{c} {1\over \Box}\left(  -(\partial_{\mu}h^{\mu\nu})\partial_{\nu}-h^{\mu\nu}\partial_{\mu}\partial_{\nu}+{1\over 2}((\partial_{\rho} h)\partial^{\rho}+h\Box)+2\xi(\Box h-\partial_{\mu}\partial_{\nu} h^{\mu\nu})
 \right) \right]\nonumber\\
 &=&-\lambda^2   \int {d^6p\over(2\pi)^{6}}{d^6 p' \over(2\pi)^{6}}\tilde{\phi}_{c}(p')\tilde{\phi}_{c}(-p)\tilde{h}_{\mu\nu}(k)  I^{\mu\nu}(p,p'),\label{2e}
 \end{eqnarray}
where again $k^{\mu}=p^{\mu}-p'^{\mu}$ and $I^{\mu\nu}(p,p')$ is defined by
\begin{eqnarray}
I^{\mu\nu}(p,p')&=&\int {d^6l \over (2\pi)^{6}} {1\over l^2(l+p)^2 (l+p')^2}[ k^{\mu}(l^{\nu}+p'^{\nu})+(l^{\mu}+p'^{\mu})(l^{\nu}+p'^{\nu})\nonumber\\
&\,\,&-{\eta^{\mu\nu}\over 2} \left( k\cdot(l+p')+(l+p')^2  \right)+2\xi(k^{\mu}k^{\nu}-k^2\eta^{\mu\nu})   ].
\end{eqnarray} 
After introducing Feynman parameters to combine the denominators, shifting the integration variables, and performing the momentum integral, we obtain
\begin{eqnarray}
I^{\mu\nu}(p,p')&=& 2\int_{0}^{1}dy \int_{0}^{1-y}\int {d^{d} l' \over (2\pi)^{d}}{1\over (l'+xyk^2)^3}\{ x(x-1)p^{\mu}p^{\nu}+y(y-1)p'^{\mu}p'^{\nu}\nonumber\\
& &+ x y p'^{\mu}p^{\nu}+(1-y)(1-x)p^{\mu}p'^{\nu}+{k^2\over 4}\eta^{\mu\nu}[(1-y)(1-x)+xy ]\nonumber\\
&&+2\xi (k^{\mu}k^{\nu}-k^2\eta^{\mu\nu})+({1\over d}-{1\over 2})l'^{2} \eta^{\mu\nu}\}\nonumber\\
&=& {-i\over(4\pi)^{d/2}}\Gamma(3-{d\over 2})(-k^2)^{{d\over 2}-3} [ -{\Gamma({d\over 2}-2)\Gamma({d\over 2})\over \Gamma(d-1)}(p^{\mu}p^{\nu}+p'^{\mu}p'^{\nu})+{\Gamma({d\over 2}-1)^{2} \over \Gamma(d-1)}p'^{\mu}p^{\nu}\nonumber\\
&&+\left( {\Gamma({d\over 2}-1)^{2} \over \Gamma(d-1)}+{\Gamma({d\over 2}-2)^{2} \over \Gamma(d-2)}\right)p^{\mu}p'^{\nu}-{\Gamma({d\over 2}-2)\Gamma({d\over2})\over 4\Gamma(d-2)}\left( {1\over({d\over 2}-1)^2} - {1\over {d\over 2}-2} \right)k^2\eta^{\mu\nu}\nonumber\\
&&+2\xi{\Gamma({d\over 2}-2)^2 \over \Gamma(d-3)}(k^{\mu}k^{\nu}-k^2\eta^{\mu\nu}) ] -{i\over 2 (4\pi)^{d/2}}{\Gamma(2-{d\over 2})\Gamma({d\over 2}-1)\Gamma({d\over 2})\over \Gamma(d-1)}(-k^2)^{{d\over 2}-2}\eta^{\mu\nu}\nonumber\\
&=& {i\over (4\pi)^3}[ \left({1\over12}-\xi \right) \left({1\over \bar{\epsilon} }-{\rm ln}\,(-k^2) \right)(p^{\mu}p^{\nu}+p'^{\mu}p'^{\nu}) \nonumber\\
&\,\,\,\,\,\,\,\,\,\,\,\,\,\,&-\left({1\over8}-\xi \right) \left({1\over \bar{\epsilon} }-{\rm ln}\,(-k^2) \right)(p'^{\mu}p^{\nu}+p^{\mu}p'^{\nu})\nonumber\\
&\,\,\,\,\,\,\,\,\,\,\,\,\,\,&+\left({5\over24}-2 \xi \right) \left({1\over \bar{\epsilon} }-{\rm ln}\,(-k^2) \right)p\cdot p' \eta^{\mu\nu} +{\rm finite} \,\,].\label{2eint}
\end{eqnarray}
In the derivation of this result, we have used the on-shell condition $p^2=p'^2=0$ and $p\cdot p' =-{k^2\over 2}$. After renormalization in modified minimal subtraction scheme$\footnote{For details on the renormalization of $\phi^3$ theory in curved spacetime, see \cite{Jack1, Kodaria}.}$, the results~(\ref{2e}) and~(\ref{2eint}) allow us to write down the leading order non-local correction to the on-shell effective action. We will denote this correction by $\Gamma^{(1)}_{\xi,{\rm nonlocal}}$. In position space, it reads
\begin{equation}
\Gamma^{(1)}_{\xi,{\rm nonlocal}}=-{1\over 2} \int d^6 x \,h^{\mu\nu}(x)\tilde{T}_{\mu\nu}(x), \label{nonlocaleff}
\end{equation}
where
\begin{eqnarray}
\tilde{T}_{\mu\nu}(x)&=&-{\lambda^2\over (4\pi)^3} \,{\rm ln}\left( {\Box \over \mu^2} \right)[ \left( {1\over 6}-2\xi \right)\phi (x)\partial_{\mu}\partial_{\nu} \phi(x) +\left( {1\over 4}-2\xi \right)\partial_{\mu} \phi(x)\partial_{\nu}\phi(x)\nonumber\\
&&-\left( {5\over 24}-2\xi \right)\eta_{\mu\nu} \partial_{\rho} \phi(x) \partial^{\rho} \phi(x)] \label{nonlocalT}
\end{eqnarray}
is the non-local correction to the energy-momentum tensor.
 
 With our results~(\ref{nonlocaleff}) and~(\ref{nonlocalT}), we will now consider two special cases by choosing two different values of the parameter $\xi$. For $\xi=0$ where the scalar $\phi$ is minimally coupled to gravity, the induced non-local contribution to the energy-momentum tensor takes the form
\begin{eqnarray}
\tilde{T}_{\mu\nu}(x)=-{\lambda^2\over 2(4\pi)^3} \,{\rm ln}\left( {\Box \over \mu^2} \right)\left( {1\over 3}\phi (x)\partial_{\mu}\partial_{\nu} \phi(x) +{1\over 2}\partial_{\mu} \phi(x)\partial_{\nu}\phi(x)
-{5\over 12}\eta_{\mu\nu} \partial_{\rho} \phi(x) \partial^{\rho} \phi(x)\right),
\end{eqnarray}
from which it follows that
\begin{equation}
\tilde{T}_{\mu}^{\mu}(x)={\lambda^2\over 2(4\pi)^3}\Box\,{\rm ln}\left( {\Box \over \mu^2} \right)\phi^2(x).
\end{equation}
To obtain this result, we have used the integration by parts and the on-shell condition. It is not difficult to see that this in fact corresponds to the non-local quantum correction to the virial current. The modification of the trace of the energy-momentum tensor from non-local quantum effect is also observed in \cite{Donoghue} for the minimally coupled massless scalar QED. 

For the classically Weyl invariant case with $\xi={1\over 10}$, the correction~(\ref{nonlocalT}) reads
\begin{eqnarray}
\tilde{T}_{\mu\nu}(x)=-{\lambda^2\over 60 (4\pi)^3} \,{\rm ln}\left( {\Box \over \mu^2} \right)\left(3\partial_{\mu} \phi(x)\partial_{\nu}\phi(x)-2\phi (x)\partial_{\mu}\partial_{\nu} \phi(x) 
-{1\over 2}\eta_{\mu\nu} \partial_{\rho} \phi(x) \partial^{\rho} \phi(x)\right). \label{WeylT}
\end{eqnarray}
Note that terms in parenthesis in~(\ref{WeylT}) is proportional to the order-$\lambda^0$ part of the classical energy-momentum tensor~(\ref{clT}) at $\xi={1\over10}$, after taking the one-shell condition. Thus the non-local correction~(\ref{WeylT}) is traceless on-shell and only contributes to the symmetric traceless part of the energy-momentum tensor. The fact that this order-$\lambda^2$ result does not modify the trace of the energy-momentum tensor is consistent with the expression of the trace anomaly shown explicitly in~(\ref{traceanomaly}), which says that the trace anomaly of the energy-momentum tensor is of order $\lambda^3$.

\section{Conclusion}
Although $\phi^3$ theory is not an interesting theory from the phenomenological point of view, it serves as an intriguing theoretical model for one to probe and gain a deeper insight into the nature of quantum field theory. In this paper, we studied massless $\phi^3$ theory in six-dimensional curved spacetime and used the background field method to construct the non-local quantum correction to the on-shell effective action. These non-local corrections are induced by massless degrees of freedom propagating long distances in loops. We show by explicit evaluation in perturbation theory that the lowest-order non-local correction is of order $\lambda^2$. As an application, the resulting effective action is used to obtain the non-local contribution to the energy-momentum tensor. Our results are expressed in~(\ref{nonlocaleff}) and~(\ref{nonlocalT}).

One key observation from our results is that for a scalar minimally coupled to gravity, the trace of the energy-momentum tensor is corrected via the modification of the virial current by this non-local effect. On the contrary, if the scalar couples to gravity in a classically Weyl invariant way, then only the traceless part of the energy-momentum tensor is affected at this order. The implications of these results remain to be explored.

\begin{acknowledgments}
This research was supported in part by the National Nature Science Foundation of China under Grant No. 10805024 and the 555 talent project of Jiangxi Province.
\end{acknowledgments}

\appendix
\section{Useful Integrals}
Here we list explicit forms for the massless bubble and triangle integrals in $d=6-\epsilon$ dimensional flat space needed to derive the trace anomaly in Sec. II. For calculational details leading to these expressions, see \cite{Davydychev, Campbell}.
\begin{eqnarray}
I_{2}(k)&=&\int {d^d l \over (2\pi)^d } {1\over l^2 (k+l)^2}={i\over (4\pi)^{d/2}}\Gamma(2-{d\over2}){\Gamma({d\over 2}-1)^2\over \Gamma(d-2)}(-k^2)^{{d\over 2}-2}\nonumber\\
&=& {i\over 6(4\pi)^3}k^2\left( {2\over \epsilon} -\gamma+{\rm ln}\, 4\pi +{8\over 3} -{\rm ln}\,(-k^2) +{\rm O}(\epsilon)\right).\label{buint}
\end{eqnarray}
\begin{eqnarray}
I_{3}(p_1,p_2)&=&\int {d^d l \over (2\pi)^d } {1\over l^2 (l+p_1)^2(l-p_2)^2}\nonumber\\
&=&{i\over 2(4\pi)^{3-\epsilon/2}}\left(p_{1}^{2}I_3[x_1]+p_{2}^{2}I_{3}[x_2]-{\Gamma(1-{\epsilon\over 2})^2 \Gamma(1+{\epsilon\over 2})\over \Gamma(1-\epsilon)}\left( {2(-s_{12})^{-\epsilon/2}\over \epsilon}+3 \right)  \right).\label{tri}
\end{eqnarray} 
Here we have defined 
\begin{equation}
I_{3}[x_{1}]={1\over \triangle}\left( p_{2}^{2}(s_{12}+p_{1}^{2}-p_{2}^{2})I_{3}[1]+(p_{1}^{2}+p_{2}^{2}-s_{12})\,{\rm ln}\left({s_{12}\over p_{1}^{2} }\right)-2p_{2}^{2}\, {\rm ln}\left({s_{12}\over p_{2}^{2}}\right)   \right),\label{finite1}
\end{equation}
\begin{equation}
I_{3}[1]={1\over \sqrt{-\triangle}}\left( {\rm ln}(a^{+}a^{-})\,{\rm ln}\left( {1+a^{+} \over 1-a^{-}} \right)+ 2{\rm Li_{2}}(a^{+})-2{\rm Li_{2}}(a^{-}) \right)\label{finite2}
\end{equation}
with ${\rm Li_{2}}$ being the dilogarithm function, $a^{\pm}={s_{12}+p_{2}^{2}-p_{1}^{2} \pm \sqrt{-\triangle} \over 2s_{12}}$, and $\triangle=-p_{1}^{4}-p_{2}^{4}-s_{12}^{2}+2p_{1}^{2}p_{2}^{2}+2p_{1}^{2}s_{12}+2p_{2}^{2}s_{12}$. $I_{3}[x_{2}]$ is obtained from $I_{3}[x_{1}]$ by exchanging $p_1$ and $p_2$.



\end{document}